\documentclass[aps,onecolumn,showpacs,nofootinbib,showkeys]{revtex4-2}
\usepackage[margin=2.0cm]{geometry}

\RequirePackage[T1]{fontenc}

\usepackage{epsfig,graphicx,amsmath,amssymb,bm}
\RequirePackage{mathptmx}      
\usepackage{mathrsfs}
\usepackage{amssymb}
\RequirePackage{color}
\usepackage{changes}
\usepackage{slashed}
\usepackage{multirow}
\usepackage{scalerel}
\usepackage{tikz-feynman}
\usepackage{textcomp}
\usepackage{subcaption}
\usepackage[english]{babel}
\usepackage{float}
\RequirePackage{hyperref}
\hypersetup{
    linktocpage,
    colorlinks,
    citecolor=blue,
    filecolor=black,
    linkcolor=blue,
    urlcolor=blue,
}
\usepackage{nccmath}

\begin{document}

\title{Description of the processes $e^+e^- \to K^+K^-\pi^0$ and $e^+e^- \to K^+K^-\eta$ within the extended NJL model
}


\author{M.K. Volkov$^{1}$}\email{volkov@theor.jinr.ru}
\author{A.A. Pivovarov$^{1}$}\email{pivovarov@theor.jinr.ru}
\author{K. Nurlan$^{1,2,3}$}\email{nurlan@theor.jinr.ru}

\affiliation{$^1$ Bogoliubov Laboratory of Theoretical Physics, JINR, 
                 141980 Dubna, Moscow region, Russia \\
                $^2$ The Institute of Nuclear Physics, Almaty, 050032, Kazakhstan \\
                $^3$ L.N. Gumilyov Eurasian National University, Astana, 010008, Kazakhstan}   


\begin{abstract}
In the framework of the $U(3) \times U(3)$ extended Nambu--Jona-Lasinio model, the processes $e^+e^- \to K^+K^-\pi^0$ and $e^+e^- \to K^+K^-\eta$ are described taking into account both ground and first radially excited intermediate meson states. It is shown that channels with radially excited $\phi(1680)$ meson are dominant in both processes.
The influence on the results of the appearance of phase factors of radially excited intermediate states, the existence of which is indicated by experimental data, is discussed.


\end{abstract}


\maketitle


\section{\label{Intro}Introduction}
Low-energy processes of $KK\pi$ and $KK\eta$ meson production in $e^+e^-$ annihilation are the subject of intensive research in recent experiments at the electron-positron colliders PEP-II, VEPP-2000, BEPC II by the BaBar,   \cite {BaBar:2007ceh}, SND \cite{Achasov:2018ygm,SND:2020qmb}, CMD-3 \cite{Ivanov:2019crp} and BES III \cite{BESIII:2022wxz} collaborations. The study of these processes can be useful for calculating the effect of hadronic vacuum polarization and subsequently for estimating the contribution of the latter to the muon anomalous magnetic moment ~\cite{Davier:2017zfy}.
Indeed, the $e^+e^- \to K \Bar{K} \pi$ process gives a significant contribution to the total cross section for electron-positron annihilation into hadrons (about 12\% at $E_{c.m.}=1.65$ GeV) \cite{Achasov:2018ygm,SND:2020qmb}. In addition, it is important to study more deeply the internal mechanisms of meson interactions and determine the parameters of intermediate meson states. Experimental data on the processes $e^+e^- \to K^+K^-[\pi^0,\eta]$ in both cases indicate the dominant role of intermediate channels with vector mesons $\phi$ and $\phi'$.


Early experimental data on the cross section of $e^+e^- \to K^+ K^- \pi^0$ were obtained by the DM 2 collaboration \cite{Bisello:1991kd}. More accurate measurements of the cross section were obtained in the latest experiments by BaBar \cite{BaBar:2007ceh}, SND \cite{Achasov:2018ygm,SND:2020qmb} and CMD-3 \cite{Ivanov:2019crp}. Experiments measuring total cross sections indicate that the process $e^+e^- \to K^+ K^- \pi^0$ goes through intermediate channels $K^*(892)K$, $\phi(1020 )\pi$, $K^*_2(1430)K$.
However, in the energy range $<2$ GeV the contribution of intermediate channels with tensor mesons $K^*_2(1430)K$ turns out to be negligible \cite{SND:2020qmb}. In turn, the $\phi(1020)\pi$ channel is suppressed according to the Okubo-Zweig-Iizuki (OZI) rule and can only occur due to the mixing of the isoscalar states $\omega$ and $\phi$. The process $e^+e^- \to K^+ K^- \eta$ mainly goes through an intermediate channel containing the decay $\phi'\to\phi\eta$ and, accordingly, is a convenient process for determining parameters of the first radially excited meson $\phi'$  \cite{BaBar:2007ceh,Achasov:2018ygm,Ivanov:2019crp}.


In the processes $e^+e^- \to K^+K^-\pi^0$ and $e^+e^- \to K^+K^-\eta$, the threshold for the production of final mesons is significantly higher than the masses of intermediate mesons in ground states. Therefore, the first radial excitations of mesons play a main role in these processes.

To theoretically study these processes in the energy region below 2 GeV, it is impossible to use QCD perturbation theory. Therefore, it is necessary to apply different phenomenological models.
In a recent paper \cite{Qin:2024ulb}, the above processes were studied within resonance chiral theory. There the amplitudes of processes were fixed by fitting to the experimental cross section and the invariant mass spectrum.

One of the models that successfully describes the low-energy interaction processes of four meson nonets (scalar, pseudoscalar, vector and axail-vector) in the ground state is the Nambu-Jona-Lasinio (NJL) model~\cite{Volkov:1986zb, Ebert:1985kz, Vogl:1991qt, Klevansky:1992qe, Ebert:1994mf, Volkov:2005kw, Buballa:2003qv}.
A distinctive feature of our version of the $U(3) \times U(3)$ chiral NJL model \cite{Volkov:1986zb, Ebert:1985kz,Ebert:1994mf, Volkov:2005kw} from other versions of the model ~\cite{Vogl:1991qt, Klevansky:1992qe} is a relatively large value of the four-dimensional cutoff parameter $\Lambda_4 =1260$ MeV. This value is comparable in magnitude with the masses of the first radially excited states of four meson nonets.
This gives grounds for an attempt to include mesons in the first radially excited states within the model without significant chiral symmetry violation. A similar model was proposed by one of the authors of the current work ~\cite{Volkov:1996br, Volkov:1996fk}.
This model allows one to describe low-energy meson interactions at energies not exceeding 2 GeV based on the chiral symmetry of strong interactions \cite{Volkov:1996br,Volkov:1996fk,Volkov:2005kw, Volkov:1999yi, Volkov:2017arr}.
In this model, the first radial excitations of mesons are introduced using a polynomial-type form factor of the lowest order in the transverse relative momentum of quarks in the meson $f(k_{\perp}) = c\left(1 + d k_{\perp}^2\right)$, where $c$ is a constant involved in determining the meson masses, $d$ being a slope parameter determined from the requirement that the introduction of excited states does not change the vacuum condensate and therefore the quark masses. The introduction of radially excited meson states into the model leads to the appearance of nondiagonal terms that describe the possibility of transitions between ground and excited states in the free meson Lagrangian. Its diagonalization is carried out by redefining the fields by introducing mixing angles \cite{Volkov:1999yi, Volkov:2017arr}.

In this work, the description of the processes $e^+e^- \to K^+ K^- \pi^0$ and $e^+e^- \to K^+ K^- \eta$ within the extended NJL model is presented. Attention is also paid to discussing the appearance of phase factors of the excited states of mesons the existence of which is indicated by the above experiments is also paid.


\section{Lagrangian of the NJL model}
The interaction Lagrangian of the extended NJL model containing the vertices necessary for studying the processes under consideration takes the form~\cite{Volkov:1999yi,Volkov:2005kw,Volkov:2017arr}:
\begin{eqnarray}
	\label{Lagrangian}
		\Delta L_{int} & = &
		\bar{q} \biggl[ 
		i \gamma^{5} \lambda_{0}^{\pi} \left(a_{\pi}{\pi}^{0} + b_{\pi}{\pi'}^{0}\right) + i \gamma^{5} \sum_{j = \pm} \lambda_{j}^{K} \left(a_{K}{K}^{j} + b_{K}{K'}^{j}\right) + \frac{1}{2} \sum_{j = \pm} \lambda_{j}^{K^*} \left(a_{K^*}{K^*}^{j} + b_{K^*}{K^{*}}'^{j}\right) \nonumber \\
		&& 
		+ \frac{1}{2} \gamma^{\mu} \lambda_{0}^{\rho} \left(a_{\rho}\rho^{0}_{\mu} + b_{\rho}\rho'^{0}_{\mu} \right) + \frac{1}{2} \gamma^{\mu} \lambda^{\omega} \left(a_{\rho}\omega_{\mu} + b_{\rho}\omega'_{\mu} \right) + \frac{1}{2} \gamma^{\mu} \lambda^{\phi} \left(a_{\phi}\phi_{\mu} + b_{\phi}\phi'_{\mu} \right)	
		+ i\gamma_{5} \sum_{i = u, s} \lambda_{i} a^{i}_{\eta}\eta
		\biggl]q,
\end{eqnarray}
where $q$ and $\bar{q}$ are the of $u, d$ and $s$ quark triplets, $\lambda$ are linear combinations of Gell-Mann matrices and factors $a_{M}$ and $b_ {M}$ arise as a result of diagonalization of the Lagrangian and take the following form:
\begin{eqnarray}
	a_{M} = \frac{1}{\sin(2\theta_{M}^{0})}\left[g_{M}\sin(\theta_{M} + \theta_{M}^{0}) +
	g'_{M}f_{M}(k_{\perp}^{2})\sin(\theta_{M} - \theta_{M}^{0})\right], \nonumber\\
	b_{M} = \frac{-1}{\sin(2\theta_{M}^{0})}\left[g_{M}\cos(\theta_{M} + \theta_{M}^{0}) +
	g'_{M}f_{M}(k_{\perp}^{2})\cos(\theta_{M} - \theta_{M}^{0})\right],
\end{eqnarray}
where $M$ denotes the corresponding meson. The mixing angles of the ground and first radially excited meson states are given in Table~\ref{tab_mixing}.
\begin{table}[h!]
\begin{center}
\begin{tabular}{ccccccc}
\hline
   & $\pi$ & $K$ & $K^*$ & $\rho$ & $\omega$ & $\phi$ \\
\hline
$\theta_M$ & $59.48^{\circ}$	& $58.11^{\circ}$ & $84.74^{\circ}$ &  $81.80^{\circ}$  & $81.80^{\circ}$ & $68.4^{\circ}$  \\
$\theta^0_M$ & $59.12^{\circ}$ & $55.52^{\circ}$ & $59.56^{\circ}$ & $61.50^{\circ}$  & $61.50^{\circ}$ & $57.13^{\circ}$  \\
\hline
\end{tabular}
\end{center}
\caption{The values of meson mixing angles~\cite{Volkov:2017arr}.}
\label{tab_mixing}
\end{table}

Since in the case of an $\eta$ meson four states are mixed, the corresponding factors have a slightly different structure~\cite{Volkov:2017arr}:
\begin{eqnarray}
    a^{u}_{\eta} & = & 0.71 g_{\eta^{u}} + 0.11 g'_{\eta^{u}} f_{uu}(k_{\perp}^{2}), \nonumber\\
    a^{s}_{\eta} & = & 0.62 g_{\eta^{s}} + 0.06 g'_{\eta^{s}} f_{ss}(k_{\perp}^{2}).
\end{eqnarray}

The quark-meson interaction constants that appear as a result of renormalization of the original nondiagonal Lagrangian have the following structure~\cite{Volkov:2017arr}:
\begin{eqnarray}
 g_{\pi} = g_{\eta^{u}}=\left(\frac{4}{Z_{\pi}}I_{20}\right)^{-1/2}, &\quad&
 g'_{\pi}=g'_{\eta^{u}} =  \left(4 I_{20}^{f^{2}}\right)^{-1/2}, \nonumber\\
 g_{\eta^{s}}=\left(\frac{4}{Z_{\eta^s}}I_{02}\right)^{-1/2}, &\quad&
 g'_{\eta^{s}} =  \left(4 I_{02}^{f^{2}}\right)^{-1/2}, \nonumber\\
 g_{K} =\left(\frac{4}{Z_K}I_{11}\right)^{-1/2}, &\quad&
 g'_{K} =\left(4I^{f^2}_{11}\right)^{-1/2},  \nonumber\\
 g_{K^*} =\left(\frac{2}{3}I_{11}\right)^{-1/2}, &\quad&
 g'_{K^*} =\left(\frac{2}{3}I_{11}^{f^{2}}\right)^{-1/2},  \nonumber\\
 g_{\rho} = g_{\omega} = \left(\frac{2}{3}I_{20}\right)^{-1/2}, &\quad&
 g'_{\rho} = g'_{\omega} = \left(\frac{2}{3}I_{20}^{f^{2}}\right)^{-1/2}, \nonumber\\
 g_{\phi} = \left(\frac{2}{3}I_{02}\right)^{-1/2}, &\quad&
 g'_{\phi} = \left(\frac{2}{3}I_{02}^{f^{2}}\right)^{-1/2},
\end{eqnarray}
where $Z_{\pi}$, $Z_K$ and $Z_{\eta^{s}}$ are additional renormalization constants that arise when taking into account transitions between axial-vector and pseudoscalar mesons
\begin{eqnarray}
    Z_{\pi} = \left(1 - 6\frac{m_{u}^{2}}{M^{2}_{a_{1}}}\right)^{-1}, \quad Z_{K} = \left(1 - \frac{3}{2}\frac{\left(m_{u} + m_s\right)^{2}}{M^{2}_{K_{1A}}}\right)^{-1}, \quad  Z_{\eta^s} = \left(1 - 6\frac{m_{s}^{2}}{M^{2}_{f_{1}}}\right)^{-1},
\end{eqnarray}
where $m_u = m_d = 270$~MeV, $m_s = 420$~MeV are constituent quark masses; $M_{a_{1}}$ and $M_{f_{1}}$ are masses of the corresponding axial-vector mesons; $M_{K_{1A}}$ is the effective mass as a result of mixing the states $K_{1}(1270)$ and $K_{1}(1400)$
\begin{eqnarray}
\label{MK1A}
    M^{2}_{K_{1A}} = \left(\frac{\sin^{2}{\alpha}}{M^{2}_{K_{1}(1270)}} + \frac{\cos^{2}{\alpha}}{M^{2}_{K_{1}(1400)}}\right)^{-1},
\end{eqnarray}
where the mixing angle $\alpha = 57^{\circ}$~\cite{Volkov:2019awd}.

Integrals in the definitions of the coupling constants take the form
\begin{eqnarray}
\label{integral_1}
	I_{n_{1}n_{2}}^{f^{m}} =
	-i\frac{N_{c}}{(2\pi)^{4}}\int\frac{f^{m}(k^2_{\perp})}{(m_{u}^{2} - k^2)^{n_{1}}(m_{s}^{2} - k^2)^{n_{2}}}\Theta(\Lambda_{3}^{2} - k^2_{\perp})
	\mathrm{d}^{4}k,
\end{eqnarray}
where $\Lambda_3=1030$ MeV is the three-dimensional cutoff parameter \cite{Volkov:2005kw,Volkov:2017arr}.

\section{The amplitude of the process $e^+e^- \to K^+ K^- \pi^0$}
\begin{figure*}[t]
 \centering
   \centering
   \begin{tikzpicture}
    \begin{feynman}
      \vertex (a) {\(e^+\)};
      \vertex [dot, below right=1.8cm of a] (b){};
      \vertex [below left=1.8cm of b] (c) {\(e^-\)};
      \vertex [blob, right=1.4cm of b] (d) {};
      \vertex [above right=1.8cm of d] (e) {\(K^+\)};
      \vertex [below right=1.8cm of d] (h) {\(\pi^0 (\eta)\)};
      \vertex [right=1.5cm of d] (f) {\(K^-\)};
      \diagram* {
         (a) -- [fermion] (b),
         (b) -- [fermion] (c),
         (b) -- [boson, edge label'=\({\gamma^*}\)] (d),    
         (d) -- [double] (e),  
         (d) -- [double] (h),
         (d) -- [double] (f),
      };
     \end{feynman}
    \end{tikzpicture}
   \caption{The Feynman diagram with an intermediate photon. The dashed circle represents the sum of two sub-diagrams given in Figs \ref{diagrambox}, \ref{diagram3} and \ref{diagram4}.}
 \label{diagram1}
\end{figure*}
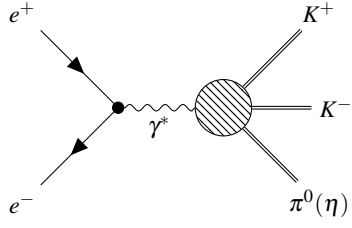

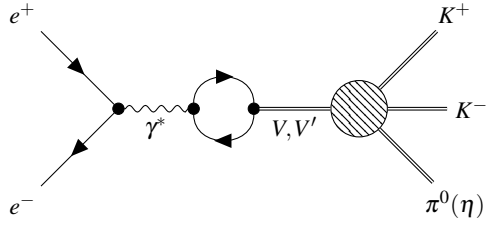
\begin{figure*}[t]
 \centering
   \centering
   \begin{tikzpicture}
    \begin{feynman}
      \vertex (a) {\(e^+\)};
      \vertex [dot, below right=1.8cm of a] (b){};
      \vertex [below left=1.8cm of b] (c) {\(e^-\)};
      \vertex [dot, right=1.0cm of b] (d) {};
      \vertex [dot, right=0.8cm of d] (l) {};
      \vertex [blob, right=1.4cm of l] (g) {};
      \vertex [above right=1.8cm of g] (e) {\(K^+\)};
      \vertex [below right=1.8cm of g] (h) {\(\pi^0(\eta)\)};
      \vertex [right=1.5cm of g] (f) {\(K^-\)};
      \diagram* {
         (a) -- [fermion] (b),
         (b) -- [fermion] (c),
         (b) -- [boson, edge label'=\({\gamma^*}\)] (d),
         (d) -- [fermion, inner sep=1pt, half left] (l),
         (l) -- [fermion, inner sep=1pt, half left] (d),
         (l) -- [double, edge label'=\({V, V'} \)] (g),
         (g) -- [double] (e),  
         (g) -- [double] (h),
         (g) -- [double] (f),
      };
     \end{feynman}
    \end{tikzpicture}
   \caption{
   The Feynman diagram with intermediate vector mesons $V=\rho, \omega, \phi$ and $V'=\rho', \omega', \phi'$. The dashed circle represents the sum of two sub-diagrams given in Figs. \ref{diagrambox}, \ref{diagram3} and \ref{diagram4}.}
 \label{diagram2}
\end{figure*}

The diagrams describing the process $e^+e^- \to K^+ K^- \pi^0$ are shown in Figures \ref{diagram1} and \ref{diagram2}.

\begin{figure*}[t]
 \centering
   \centering
   \begin{tikzpicture}
    \begin{feynman}
      \vertex [dot] (a) {};
      \vertex [dot, above right=1.8cm of a] (c){}; 
      \vertex [dot, below right=1.8cm of a] (e){};
      \vertex [dot, right=2.4cm of a] (d) {};
      \vertex [right=1.4cm of c] (f) {\(K^+\)};
      \vertex [right=1.4cm of d] (g) {\(K^-\)};
      \vertex [right=1.4cm of e] (h) {\(\pi^0(\eta)\)};
      \diagram* {
         (a) -- [fermion] (c),
         (c) -- [fermion] (d),
         (d) -- [fermion] (e),
         (e) -- [fermion] (a),
         (c) -- [double] (f),
         (d) -- [double] (g),
         (e) -- [double] (h),
      };
     \end{feynman}
    \end{tikzpicture}
   \caption{Box quark diagram describing the direct production of $V \to K^+K^- \pi^0(\eta)$.
   }
 \label{diagrambox}
\end{figure*}
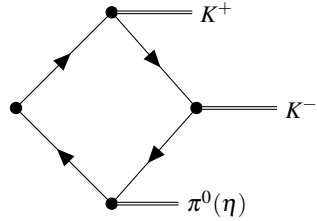%

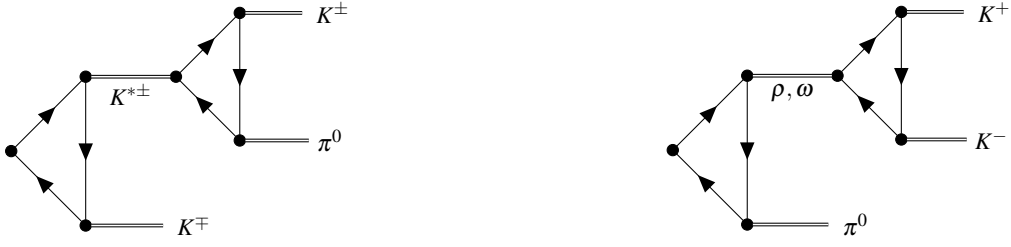
\begin{figure*}[t]
 \centering
  \begin{subfigure}{0.5\textwidth}
   \centering
   \begin{tikzpicture}
     \begin{feynman}
      \vertex [dot] (d) {};      
      \vertex [dot, above right=1.4cm of d] (e) {};
      \vertex [dot, below right=1.4cm of d] (h) {};
      \vertex [dot, right=1.2cm of e] (f) {};
      \vertex [dot, above right=1.2cm of f] (n) {};  
      \vertex [dot, below right=1.2cm of f] (m) {};   
      \vertex [right=1.2cm of n] (l) {\(\ K^{\pm} \)}; 
      \vertex [right=1.2cm of m] (s) {\(\pi^0 \)};  
      \vertex [right=1.4cm of h] (k) {\(\ K^{\mp} \)}; 
      \diagram* {         
         (d) -- [fermion] (e),  
         (e) -- [fermion] (h),
         (d) -- [anti fermion] (h),
         (e) -- [double, edge label'=\({ K^{*\pm}} \)] (f),
         (f) -- [fermion] (n),
         (n) -- [fermion] (m),
         (f) -- [anti fermion] (m), 
         (h) -- [double] (k),
         (n) -- [double] (l),
	 (m) -- [double] (s),
      };
     \end{feynman}
    \end{tikzpicture}
  \end{subfigure}%
 \centering
 \begin{subfigure}{0.5\textwidth}
  \centering
   \begin{tikzpicture}
     \begin{feynman}
      \vertex [dot] (d) {};      
      \vertex [dot, above right=1.4cm of d] (e) {};
      \vertex [dot, below right=1.4cm of d] (h) {};
      \vertex [dot, right=1.2cm of e] (f) {};
      \vertex [dot, above right=1.2cm of f] (n) {};  
      \vertex [dot, below right=1.2cm of f] (m) {};   
      \vertex [right=1.2cm of n] (l) {\(\ K^{+} \)}; 
      \vertex [right=1.2cm of m] (s) {\( K^- \)};  
      \vertex [right=1.4cm of h] (k) {\(\ \pi^0 \)}; 
      \diagram* {         
         (d) -- [fermion] (e),  
         (e) -- [fermion] (h),
         (d) -- [anti fermion] (h),
         (e) -- [double, edge label'=\({\rho, \omega} \)] (f),
         (f) -- [fermion] (n),
         (n) -- [fermion] (m),
         (f) -- [anti fermion] (m), 
         (h) -- [double] (k),
         (n) -- [double] (l),
	 (m) -- [double] (s),
      };
     \end{feynman}
    \end{tikzpicture}
  \end{subfigure}%
 \caption{The $VKK\pi$ vertex with two triangle quark loops connected by virtual vector mesons $K^{*\pm}$, $\rho$ and $\omega$.
 }
 \label{diagram3}
\end{figure*}%

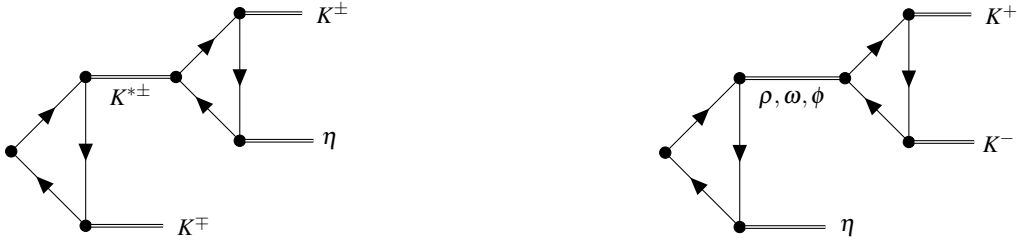
\begin{figure*}[t]
 \centering
  \begin{subfigure}{0.5\textwidth}
   \centering
   \begin{tikzpicture}
     \begin{feynman}
      \vertex [dot] (d) {};      
      \vertex [dot, above right=1.4cm of d] (e) {};
      \vertex [dot, below right=1.4cm of d] (h) {};
      \vertex [dot, right=1.2cm of e] (f) {};
      \vertex [dot, above right=1.2cm of f] (n) {};  
      \vertex [dot, below right=1.2cm of f] (m) {};   
      \vertex [right=1.2cm of n] (l) {\(\ K^{\pm} \)}; 
      \vertex [right=1.2cm of m] (s) {\(\eta \)};  
      \vertex [right=1.4cm of h] (k) {\(\ K^{\mp} \)}; 
      \diagram* {         
         (d) -- [fermion] (e),  
         (e) -- [fermion] (h),
         (d) -- [anti fermion] (h),
         (e) -- [double, edge label'=\({ K^{*\pm}} \)] (f),
         (f) -- [fermion] (n),
         (n) -- [fermion] (m),
         (f) -- [anti fermion] (m), 
         (h) -- [double] (k),
         (n) -- [double] (l),
	 (m) -- [double] (s),
      };
     \end{feynman}
    \end{tikzpicture}
  \end{subfigure}%
 \centering
 \begin{subfigure}{0.5\textwidth}
  \centering
   \begin{tikzpicture}
     \begin{feynman}
      \vertex [dot] (d) {};      
      \vertex [dot, above right=1.4cm of d] (e) {};
      \vertex [dot, below right=1.4cm of d] (h) {};
      \vertex [dot, right=1.4cm of e] (f) {};
      \vertex [dot, above right=1.2cm of f] (n) {};  
      \vertex [dot, below right=1.2cm of f] (m) {};   
      \vertex [right=1.2cm of n] (l) {\(\ K^{+} \)}; 
      \vertex [right=1.2cm of m] (s) {\( K^- \)};  
      \vertex [right=1.4cm of h] (k) {\(\ \eta \)}; 
      \diagram* {         
         (d) -- [fermion] (e),  
         (e) -- [fermion] (h),
         (d) -- [anti fermion] (h),
         (e) -- [double, edge label'=\({\rho, \omega, \phi} \)] (f),
         (f) -- [fermion] (n),
         (n) -- [fermion] (m),
         (f) -- [anti fermion] (m), 
         (h) -- [double] (k),
         (n) -- [double] (l),
	 (m) -- [double] (s),
      };
     \end{feynman}
    \end{tikzpicture}
  \end{subfigure}%
 \caption{The $VKK\pi$ vertex with two triangle quark loops connected by virtual vector mesons $K^{*\pm}$,  $\rho$, $\omega$ and $\phi$.
 }
 \label{diagram4}
\end{figure*}%

The amplitude of this process in the extended NJL model takes the following form:
\begin{eqnarray}
    \mathcal{M} = \frac{16 \pi \alpha_{em}}{q^2}L_{\mu} \mathcal{H}^{\mu},
    \label{amplitude_ee}
\end{eqnarray}
where $q$ is the momentum of the first intermediate state; $L_{\mu}=\bar{e}\gamma_\mu e$ is the lepton current. The hadronic part of the amplitude takes the form
\begin{eqnarray}
    \mathcal{H}^{\mu} = \left[\mathcal{M}_{K^{*+}} + \mathcal{M}_{K^{*-}} + \mathcal{M}_{\omega} + \mathcal{M}_{\rho} + \mathcal{M}_{K^{*'+}} + \mathcal{M}_{K^{*'-}} + \mathcal{M}_{\omega'} + \mathcal{M}_{\rho'} + \mathcal{M}_{box}\right] e^{\mu \nu \lambda \delta} p_{K^+ \nu} p_{K^- \lambda} p_{\pi^0 \delta}.
\end{eqnarray}
Here $p_{K^+}$, $p_{K^-}$ and $p_{\pi^0}$ are the momenta of mesons in the final states;
$\mathcal{M}_{K^{*\pm}}$, $\mathcal{M}_{\omega}$, $\mathcal{M}_{\rho}$, $\mathcal{M}_{K^{*'\pm}}$, $\mathcal{M}_{\omega'}$ and $\mathcal{M}_{\rho'}$ are
the contributions from the diagrams with $K^*$, $\omega$ and $\rho$ mesons and their first radial excitations as second intermediate states;
$\mathcal{M}_{box}$ is the contribution from the box diagram
\begin{eqnarray}
    \mathcal{M}_{K^{*(\prime)\pm}} & = & \mathcal{M}_{cK^{*(\prime)\pm}} + \mathcal{M}_{\rho K^{*(\prime)\pm}} + \mathcal{M}_{\omega K^{*(\prime)\pm}} + \mathcal{M}_{\phi K^{*(\prime)\pm}} + \mathcal{M}_{\rho' K^{*(\prime)\pm}} + \mathcal{M}_{\omega' K^{*(\prime)\pm}} + \mathcal{M}_{\phi' K^{*(\prime)\pm}}, \nonumber\\
    \mathcal{M}_{\omega^{(\prime)}} & = & \mathcal{M}_{c\omega^{(\prime)}} + \mathcal{M}_{\rho\omega^{(\prime)}} + \mathcal{M}_{\rho'\omega^{(\prime)}}, \nonumber\\
    \mathcal{M}_{\rho^{(\prime)}} & = & \mathcal{M}_{c\rho^{(\prime)}} + \mathcal{M}_{\omega\rho^{(\prime)}} + \mathcal{M}_{\omega'\rho^{(\prime)}}, \nonumber\\
    \mathcal{M}_{box} & = & \mathcal{M}_{cbox} + \mathcal{M}_{\rho box} + \mathcal{M}_{\omega box} + \mathcal{M}_{\phi box} + \mathcal{M}_{\rho' box} + \mathcal{M}_{\omega' box} + \mathcal{M}_{\phi' box},
\end{eqnarray}
where $\mathcal{M}_{cK^{*(\prime)\pm}}$, $\mathcal{M}_{c\omega^{(\prime)}}$, $\mathcal{M}_{c\rho^{(\prime)}}$ and $\mathcal{M}_{cbox}$ are the contact contributions; others are the contributions containing the vector meson as the first intermediate state. The explicit form of these terms is presented in Appendix~\ref{app_1}.

\section{The process $e^+e^- \to K^+ K^- \pi^0$ $e^+e^- \to K^+ K^- \eta$}

Next consider the process $e^+e^- \to K^+ K^- \eta$ within the extended NJL model. Note that in this process involving the $\eta$ meson, it is necessary to take into account the mixing of light $u$ and $d$ quarks with the heavier $s$ quark due to the influence of gluon anomalies. In the NJL model, this is achieved by introducing the 't Hooft interaction, which leads to the breaking of $U(3)\times U(3)$ chiral symmetry and mixing of the four states of mesons $\eta$, $\eta'$, $\eta(1295 )$ and $\eta(1475)$ \cite{Volkov:1999yi, Volkov:2017arr}. Consequently, in this case, in contrast to the process $e^+e^- \to K^+ K^- \pi^0$, additional terms appear in the amplitude that correspond to the contributions from the diagrams with the $s$ quark part of the $\eta$ meson 

The total amplitude of the process $e^+e^- \to K^+ K^- \eta$ is determined by equation (\ref{amplitude_ee}). The hadronic part of the amplitude in the extended NJL model takes the following form:
\begin{eqnarray}
&& \mathcal{H}^\mu = \biggl[\mathcal{M}_{K^{*+}} + \mathcal{M}_{K^{*-}} + \mathcal{M}_{\omega} + \mathcal{M}_{\rho} + \mathcal{M}_{\phi} + \mathcal{M}_{K^{*'+}} + \mathcal{M}_{K^{*'-}} \nonumber \\ && \qquad
 + \mathcal{M}_{\rho'} + \mathcal{M}_{\omega'}  + \mathcal{M}_{\phi'}+ \mathcal{M}_{box} \biggl] e^{\mu \nu \lambda \delta} p_{K^+ \nu} p_{K^- \lambda} p_{\eta \delta}\,,
\end{eqnarray}
where $p_{K^+}$, $p_{K^-}$ and $p_{\eta}$ are the momenta of mesons in the final states; $\mathcal{M}_{K^{*\pm}}$, $\mathcal{M}_{\omega}$, $\mathcal{M}_{\rho}$,  $\mathcal{M}_{\phi}$, $\mathcal{M}_{K^{*'\pm}}$, $\mathcal{M}_{\omega'}$, $\mathcal{M}_{\rho'}$ and $\mathcal{M}_{\phi'}$ are the contributions of the diagrams presented in Fig. \ref{diagram4} with the second intermediate states $K^*$, $\omega$, $\rho$ and $\phi$ and their first radial excitations; $\mathcal{M}_{box}$ is the contribution from the anomalous box quark diagram (Fig. \ref{diagrambox}):
\begin{eqnarray}
\mathcal{M}_{K^{*(\prime)\pm}} & = & \mathcal{M}_{cK^{*(\prime)\pm}} + \mathcal{M}_{\rho K^{*(\prime)\pm}} + \mathcal{M}_{\omega K^{*(\prime)\pm}} + \mathcal{M}_{\phi K^{*(\prime)\pm}} + \mathcal{M}_{\rho' K^{*(\prime)\pm}} + \mathcal{M}_{\omega' K^{*(\prime)\pm}} + \mathcal{M}_{\phi' K^{*(\prime)\pm}}, \nonumber\\
    \mathcal{M}_{\omega^{(\prime)}} & = & \mathcal{M}_{c\omega^{(\prime)}} + \mathcal{M}_{\omega\omega^{(\prime)}} +\mathcal{M}_{\omega'\omega^{(\prime)}}, \nonumber\\
    \mathcal{M}_{\rho^{(\prime)}} & = & \mathcal{M}_{c\rho} + \mathcal{M}_{\rho\rho^{(\prime)}} + \mathcal{M}_{\rho'\rho^{(\prime)}}, \nonumber\\
    \mathcal{M}_{\phi^{(\prime)}} & = & \mathcal{M}_{c\phi^{(\prime)}} + \mathcal{M}_{\phi\phi^{(\prime)}} + \mathcal{M}_{\phi'\phi^{(\prime)}},
    \nonumber\\
    \mathcal{M}_{box} & = & \mathcal{M}_{cbox} + \mathcal{M}_{\rho box} + \mathcal{M}_{\omega box} + \mathcal{M}_{\phi box} + \mathcal{M}_{\rho' box} + \mathcal{M}_{\omega' box} + \mathcal{M}_{\phi' box}.
\end{eqnarray}
The explicit forms of these terms are presented in Appendix~\ref{app_2}.

\section{Numerical estimations}
The cross section for the process $e^+e^- \to K^+K^-\pi$ can be calculated in the massless approximation of colliding leptons using the following formula:
\begin{eqnarray}
    \sigma(q^2) & = & \frac{1}{2^8\pi^3q^8} \int_{m_{1 min}}^{m_{1 max}} dq_{\pi K^+}^2 \int_{m_{2 min}}^{m_{2 max}} T_{1\mu\nu}(p_1, p_2) \left[q^2 g^{\mu\nu}\left(q^2g^{\lambda\delta} - q^{\lambda}q^{\delta}\right) \right. \nonumber\\
    && \left. + q^{\mu}q^{\nu}\left(-q^2g^{\lambda\delta} + 4q^{\lambda}q^{\delta}\right)\right] T_{2\lambda\delta}(p_{\pi}, p_{K^+}, p_{K^-}) dq_{\pi K^-}^2,
\end{eqnarray}
where $T_{1\mu\nu}(p_1, p_2)$ is the squared lepton current averaged over polarizations depending only on the momenta of colliding leptons:
\begin{eqnarray}
    T_{1\mu\nu}(p_1, p_2) = p_{1\mu}p_{2\nu} + p_{2\mu}p_{1\nu} - \left(p_1, p_2\right) g_{\mu\nu}.
\end{eqnarray}
The amplitude square depending only on the momenta of the final mesons is
\begin{eqnarray}
    T_{2\lambda\delta}(p_{\pi}, p_{K^+}, p_{K^-}) = \mathcal{H}_{\lambda}^*\mathcal{H}_{\delta}.
\end{eqnarray}
The integration limits that define the physical areas of the process take the following form:
\begin{eqnarray}
    &m_{1 min} = \left(M_{\pi} + M_K\right)^2,& \nonumber\\
    &m_{1 max} = \left(\sqrt{q^2} - M_K\right)^2,& \nonumber\\
    &m_{2 min} = \frac{\left(q^2 - 2 M_K^2 +  M_{\pi}^2\right)^2}{4 q_{\pi K^+}^2} - \left(\sqrt{\frac{\left(q_{\pi K^+}^2 - M_K^2 +  M_{\pi}^2\right)^2}{4 q_{\pi K^+}^2} - M_{\pi}^2} + \sqrt{\frac{\left(q^2 - q_{\pi K^+}^2 - M_K^2\right)^2}{4 q_{\pi K^+}^2} - M_{K}^2}\right)^2,& \nonumber\\
    &m_{2 max} = \frac{\left(q^2 - 2 M_K^2 +  M_{\pi}^2\right)^2}{4 q_{\pi K^+}^2} - \left(\sqrt{\frac{\left(q_{\pi K^+}^2 - M_K^2 +  M_{\pi}^2\right)^2}{4 q_{\pi K^+}^2} - M_{\pi}^2} - \sqrt{\frac{\left(q^2 - q_{\pi K^+}^2 - M_K^2\right)^2}{4 q_{\pi K^+}^2} - M_{K}^2}\right)^2.&
\end{eqnarray}

For the process $e^+e^- \to K^+K^-\eta$ the expressions are similar with the replacement of $\pi$ meson by $\eta$.

The uncertainty of numerical calculations in the model is estimated at the level of $\pm15\%$ \cite{Volkov:2005kw, Volkov:1999yi, Volkov:2017arr}. This based on statistical analysis of numerous previous calculations within the
model and is justified by a noticeable violation of chiral symmetry in the energy region under study.

\begin{figure}[h]
\center{\includegraphics[scale = 0.8]{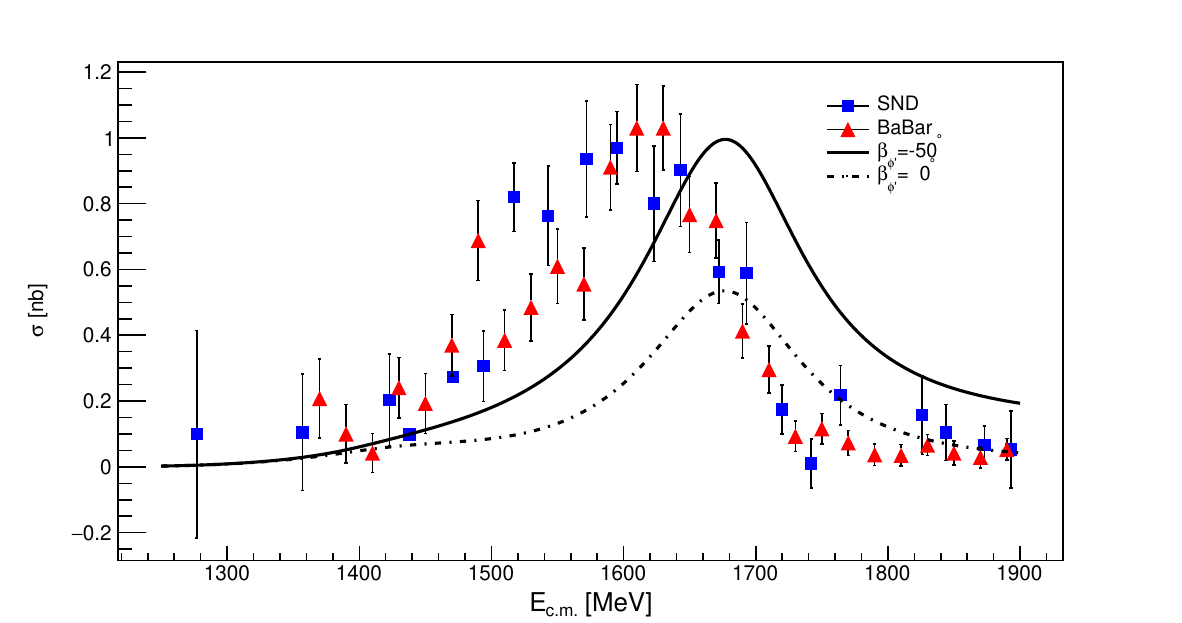}}
\caption{
Comparison of the $e^+e^- \to K^+K^-\pi^0$ process total cross section calculated in the NJL model with experimental data of the BaBar \cite{BaBar:2007ceh} and SND  \cite{SND:2020qmb} collaborations.
}
\label{KKpi}
\end{figure}

\begin{figure}[h]
\center{\includegraphics[scale = 0.8]{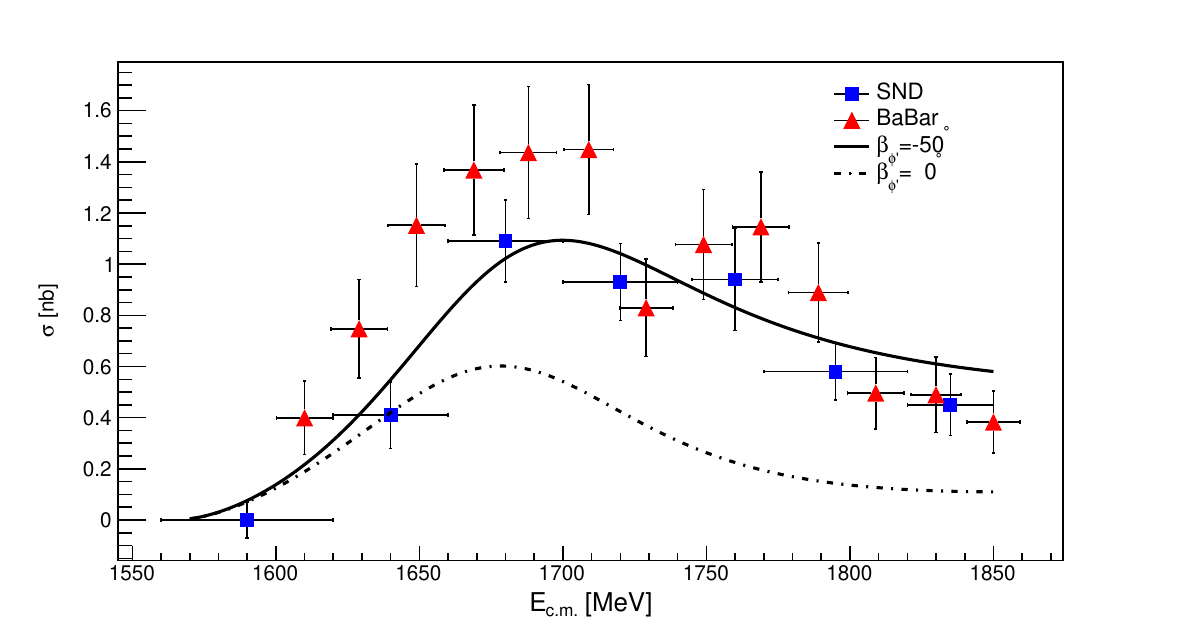}}
\caption{Comparison of the $e^+e^- \to K^+K^-\eta$ process total cross section with experimental data \cite{BaBar:2007ceh,Achasov:2018ygm}.}
\label{KKeta}
\end{figure}

The dependences of the cross section of the processes under consideration on the energy of colliding leptons obtained on the basis of the above amplitudes are presented in Figs.~\ref{KKpi} and~\ref{KKeta} by the dashed lines. 
As we can see from the graphs, in both cases the resonance height is noticeably lower than the experimental data.
Experimental results of such processes often indicate the presence of relative phases between different channels of the same process \cite{BaBar:2007ceh,SND:2020qmb,Ivanov:2019crp,Achasov:2007kw,Achasov:2013btb,Achasov:2016zvn}. The NJL model used in this work does not describe these phases. However, taking them into account can have a significant impact on the results of calculations. In the processes discussed above, the main contribution comes from the channels containing the radially excited $\phi(1680)$ meson as the first intermediate state. In the case of the $e^+e^- \to K^+K^-\eta$ process, this contribution almost completely determines the cross section since the masses of the remaining mesons are below the threshold for the production of final states.
If the phase factor $BW_{\phi'} \to e^{i\beta_{\phi'}} BW_{\phi'}$ is placed in the amplitude containing this meson, a rather strong dependence of the obtained results on $\beta_{\phi'}$. 
The dependence of the cross section of this process on the energy of colliding leptons becomes closest to the experimental values at the relative phase $\beta_{\phi'} = -50^{\circ}$. The corresponding graph is shown in Figure~\ref{KKeta} as a solid line. Taking into account phase factors before other channels of this process has virtually no effect on the result due to their insignificant contribution.


The application of the same phase factor to the process $e^+e^- \to K^+K^-\pi^0$ leads to the dependence shown in Fig.~\ref{KKpi} by the solid line. As we can see, taking into account such a phase factor for contributions containing the $\phi(1680)$ meson in this process allows one to better describe the height of the main resonance but at the same time retains some shift of this resonance relative to the experimental values.
The contributions from channels containing the first radial excited states $\rho'$ and $\omega'$ mesons play a larger role than in the $e^+e^- \to K^+K^-\eta$ process due to the lower threshold for the production of final states in this process. The cross-sectional shape of this process may depend on the phase factor of the terms describing these channels. The values $|\beta_{[\rho', \omega']}| < 30^{\circ}$ do not lead to a significant change in the shape of the curve. However, large absolute values of these angles lead to worse agreement with experimental data.

\section{Conclusion}
In this work, the processes of low-energy production of $K^+K^-\pi^0$ and $K^+K^-\eta$ mesons in electron-positron collisions are described within the extended version of the quark NJL model. The amplitudes for contact channels and channels with intermediate vector mesons in the ground and first radially excited states are considered.

The calculations show that the main contribution to the $e^+e^- \to K^+K^-\pi^0$ process total cross section comes from the channel containing the decay $\phi' \to K^{*\pm}K^\mp$. 
The channel with the $\phi’ \to \phi \eta$ decay completely determines the cross section of the process in the case of the $K^+K^-\eta$ meson production process. The channels associated with the mesons $\rho'$ and $\omega'$ mesons make a significantly smaller contribution. This especially takes place in the case of the process $e^+e^- \to K^+K^-\eta$. The remaining channels including anomalous box diagrams make a negligible contribution.
It should be noted that the results within the NJL model were obtained without using any additional arbitrary parameters, with the exception of the phase factor $\beta_{\phi'}$. 
At the same time, in the paper \cite{Qin:2024ulb} the amplitudes of the processes were fixed by fitting to the experimental cross-section using many parameters within the resonance chiral theory.

These calculations were carried out in the energy region where the chiral symmetry on which the model is based begins to be broken noticeably. In addition, vector mesons of the second and higher excitation degrees, which are not described by the extended NJL model, begin to play an important role in the energy region above $1.60$ GeV.
Therefore, within this model it would be unreasonable to expect more accurate results at the energies considered.

The obtained formulas make it possible to study the assumptions made in experimental works about the influence of phase factors in describing radially excited states. The results show that the existence of phase factors with an angle $\beta_{\phi'} = -50^\circ$ in the radially excited state $\phi(1680)$ leads to better agreement with experimental data than in the absence of such a factor. Moreover, this is true for both processes under consideration.

\appendix
\section{The explicit form of $e^+e^- \to K^+K^-\pi^0$ process channels}
\label{app_1}
The contact contributions
\begin{eqnarray}
    \mathcal{M}_{cK^{*(\prime)\pm}} & = & \frac{2}{3} \left(2 * I_{cu}^{K^{*(\prime)}K} - I_{cs}^{K^{*(\prime)}K}\right) I_{11}^{K^{*(\prime)}K\pi} \left(Z_{K^{*(\prime)}K_1} + Z_{K^{*(\prime)}a_1}\right) BW_{K^{*(\prime)\pm}}(q_{\pi K^{\pm}}^2), \nonumber\\
    \mathcal{M}_{c\omega^{(\prime)}} & = & 4 m_u I_{30}^{\omega^{(\prime)} \pi} I_{11}^{\omega^{(\prime)} K K} Z_{\omega^{(\prime)}} BW_{\omega^{(')}}(q_{KK}^2), \nonumber\\
    \mathcal{M}_{c\rho^{(\prime)}} & = & \frac{4}{3} m_u I_{30}^{\rho^{(\prime)} \pi} I_{11}^{\rho^{(\prime)} K K} Z_{\rho^{(\prime)}} BW_{\rho^{(\prime)}}(q_{KK}^2), \nonumber\\    
    \mathcal{M}_{cbox} & = & \frac{2}{3} \left(2 I_{cubox}^{KK\pi} - I_{csbox}^{KK\pi}\right),
\end{eqnarray}
where $q_{\pi K^\pm} = p_\pi + p_{K^\pm}$ and $q_{KK} = p_{K^+} + p_{K^-}$ are the momenta of the second intermediate states.

The contributions containing mesons in the first intermediate states take the form:
\begin{eqnarray}
    \mathcal{M}_{\rho^{(\prime)} K^{*(\prime)\pm}} & = & \frac{C_{\rho^{(\prime)}}}{g_{\rho}} I_{cu}^{K^{*(\prime)}K\rho^{(\prime)}} I_{11}^{K^{*(\prime)}K\pi} \left(Z_{K^{*(\prime)}K_1} + Z_{K^{*(\prime)}a_1}\right) BW_{\rho^{(\prime)}}(q^2) BW_{K^{{*(\prime)}\pm}}(q_{\pi K^{\pm}}^2), \nonumber\\
    \mathcal{M}_{\omega^{(\prime)} K^{*(\prime)\pm}} & = & \frac{1}{3} \frac{C_{\omega^{(\prime)}}}{g_{\rho}} I_{cu}^{K^{*(\prime)}K\omega^{(\prime)}} I_{11}^{K^{*(\prime)}K\pi} \left(Z_{K^{*(\prime)}K_1} + Z_{K^{*(\prime)}a_1}\right) BW_{\omega^{(\prime)}}(q^2) BW_{K^{{*(\prime)}\pm}}(q_{\pi K^{\pm}}^2), \nonumber\\
    \mathcal{M}_{\phi^{(\prime)} K^{*(\prime)\pm}} & = & -\frac{2}{3} \frac{C_{\phi^{(\prime)}}}{g_{\phi}} I_{cs}^{K^{*(\prime)}K\phi^{(\prime)}} I_{11}^{K^{*(\prime)}K\pi} \left(Z_{K^{*(\prime)}K_1} + Z_{K^{*(\prime)}a_1}\right) BW_{\phi^{(\prime)}}(q^2) BW_{K^{{*(\prime)}\pm}}(q_{\pi K^{\pm}}^2), \nonumber\\
    \mathcal{M}_{\rho^{(\prime)}\omega^{(\prime)}} & = & 4 \frac{C_{\rho^{(\prime)}}}{g_{\rho}} m_u I_{30}^{\omega^{(\prime)} \rho^{(\prime)} \pi} I_{11}^{\omega^{(\prime)} K K} Z_{\omega^{(\prime)}}  BW_{\rho^{(\prime)}}(q^2) BW_{\omega^{(\prime)}}(q_{KK}^2), \nonumber\\
    \mathcal{M}_{\omega^{(\prime)}\rho^{(\prime)}} & = & \frac{4}{3} \frac{C_{\omega^{(\prime)}}}{g_{\rho}} m_u I_{30}^{\omega^{(\prime)} \rho^{(\prime)} \pi} I_{11}^{\rho^{(\prime)} K K} Z_{\rho^{(\prime)}}  BW_{\omega^{(\prime)}}(q^2) BW_{\rho^{(\prime)}}(q_{KK}^2), \nonumber\\
    \mathcal{M}_{\rho^{(\prime)} box} & = & \frac{C_{\rho^{(\prime)}}}{g_{\rho}} I_{cubox}^{\rho^{(\prime)}KK\pi} BW_{\rho^{(\prime)}}(q^2), \nonumber\\
     \mathcal{M}_{\omega^{(\prime)} box} & = & \frac{1}{3} \frac{C_{\omega^{(\prime)}}}{g_{\rho}} I_{cubox}^{\omega^{(\prime)}KK\pi} BW_{\omega^{(\prime)}}(q^2), \nonumber\\
     \mathcal{M}_{\phi^{(\prime)} box} & = & \frac{2}{3} \frac{C_{\phi^{(\prime)}}}{g_{\phi}} I_{csbox}^{\phi^{(\prime)}KK\pi} BW_{\phi^{(\prime)}}(q^2).
\end{eqnarray}

The intermediate states are described by the Breit-Wigner propagators:
\begin{eqnarray}
    BW_{meson}(p^2) = \frac{1}{M_{meson}^2 - p^2 - i\sqrt{p^2}\Gamma_{meson}},
\end{eqnarray}
where the mesons masses and widths are taken from PDG \cite{ParticleDataGroup:2022pth}.

The constants arising when taking into account transitions between axial-vector and pseudoscalar mesons are
\begin{eqnarray}
    Z_{K^{*(\prime)}K_1} & = & 1 - 2\frac{I_{11}^{K^{*(\prime)}K_1\pi}I_{11}^{K_1K}}{I_{11}^{K^{*(\prime)}K\pi}} \frac{m_s\left(m_s + m_u\right)}{M_{K_{1A}}^2}, \nonumber\\
    Z_{K^{*(\prime)}a_1} & = & 1 - 2\frac{I_{11}^{K^{*(\prime)}Ka_1}I_{20}^{a_1\pi}}{I_{11}^{K^{*(\prime)}K\pi}} \frac{m_u\left(3m_u - m_s\right)}{M_{a_{1}}^2}, \nonumber\\
    Z_{\omega^{(\prime)}} & = & 1 - \frac{I_{11}^{\omega^{(\prime)} K_1 K}I_{11}^{K_1K}}{I_{11}^{\omega^{(\prime)} K K}} \frac{\left(m_u + m_s\right)^2}{M_{K_{1A}}^2}, \nonumber\\
    Z_{\rho^{(\prime)}} & = & 1 - \frac{I_{11}^{\rho^{(\prime)} K_1 K}I_{11}^{K_1K}}{I_{11}^{\rho^{(\prime)} K K}} \frac{\left(m_u + m_s\right)^2}{M_{K_{1A}}^2},
\label{avp_const}
\end{eqnarray}
where $M_{K_{1A}}$ is defined in (\ref{MK1A}).

In the anomalous vertices of two vector and one pseudoscalar meson interactions and the box diagram there arise convergent integrals
\begin{eqnarray}
\label{int2}
    I_{cu}^{mesons} & = & m_u \left[I_{21}^{mesons} + m_u \left(m_s - m_u\right)I_{31}^{mesons}\right], \nonumber\\
    I_{cs}^{mesons} & = & m_s \left[I_{12}^{mesons} - m_s \left(m_s - m_u\right)I_{13}^{mesons}\right], \nonumber\\
    I_{cubox}^{mesons} & = & \left(3 m_u - m_s\right)I_{31}^{mesons} + 3m_u^2\left(m_s - m_u\right)I_{41}^{mesons}, \nonumber\\
    I_{csbox}^{mesons} & = & m_s\left[I_{22}^{mesons} - m_s\left(m_s - m_u\right)I_{23}^{mesons}\right].
\end{eqnarray}
The integrals arising in quark loops with different meson vertices in the extended NJL model have the following form:
\begin{eqnarray}
    I_{n_{1}n_{2}}^{M_1, M_2, \dots, M_1', M_2', \dots} =
	-i\frac{N_{c}}{(2\pi)^{4}}\int\frac{a_{M_1} a_{M_2} \dots b_{M_1} b_{M_2} \dots}{(m_{u}^{2} - k^2)^{n_{1}}(m_{s}^{2} - k^2)^{n_{2}}}\Theta(\Lambda_{3}^{2} - k^2_{\perp})
	\mathrm{d}^{4}k,
\end{eqnarray}
where $M$ denotes the corresponding meson; $a$ and $b$ are the factors from the Lagrangian~(\ref{Lagrangian}).

The constants that arise during the transition of a photon to the first intermediate meson are determined by the mixing angles of the corresponding meson states:
\begin{eqnarray}
    C_{M} & = & \frac{1}{\sin{\left(2\theta_M^0\right)}} \left[\sin{\left(\theta_M + \theta_M^0\right)} + R\sin{\left(\theta_M - \theta_M^0\right)}\right], \nonumber\\
    C_{M'} & = & \frac{-1}{\sin{\left(2\theta_M^0\right)}} \left[\cos{\left(\theta_M + \theta_M^0\right)} + R\cos{\left(\theta_M - \theta_M^0\right)}\right]
\end{eqnarray}
where
\begin{eqnarray}
    R = \frac{I_{20}^f}{\sqrt{I_{20} I_{20}^{ff}}} &\quad& \textbf{ ($\rho$ and $\omega$ mesons cases)}, \nonumber\\
    R = \frac{I_{02}^f}{\sqrt{I_{02} I_{02}^{ff}}} &\quad& \textbf{ ( $\phi$ meson case)},
\end{eqnarray}
where the integrals $I_{20}$ and $I_{02}$ with different numbers of form factors are given in (\ref{integral_1}).

\section{The explicit form of $e^+e^- \to K^+K^-\eta$ process channels}
\label{app_2}
The contributions from the contact channels presented in Fig. \ref{diagram1} have the form
\begin{eqnarray}
    \mathcal{M}_{cK^{*\pm}} & = & \frac{2}{3} \left(2 I_{cu}^{K^*K} - I_{cs}^{K^*K}\right) \left( I_{11}^{K^*K\eta_u} + \sqrt{2} I_{11}^{K^*K\eta_s} \right) \left( Z_{K^*K_1} +1 \right) BW_{K^{*\pm}}(q_{\eta K^{\pm}}^2), \nonumber\\
    \mathcal{M}_{c\rho} & = & m_u I_{30}^{\rho \eta_u} Z_{\rho} BW_{\rho}(q_{KK}^2), \nonumber\\
    \mathcal{M}_{c\omega} & = & \frac{m_u}{3} I_{30}^{\omega \eta_u} Z_{\omega} BW_{\omega}(q_{KK}^2), \nonumber\\
    \mathcal{M}_{c\phi} & = & \frac{8\sqrt{2}m_s}{3} I_{03}^{\phi \eta_s} I_{11}^{\phi K K} Z_{\phi} BW_{\phi}(q_{KK}^2),\nonumber\\    
    \mathcal{M}_{cbox} & = & \frac{2}{3} \left(2 I_{cu1box}^{KK\eta_u} - I_{cs1box}^{KK\eta_u} - 2\sqrt{2} I_{cu2box}^{KK\eta_s} + \sqrt{2} I_{cs2box}^{KK\eta_s} \right).
\end{eqnarray}
The contributions containing mesons in the first intermediate states take the form:
\begin{eqnarray}
    \mathcal{M}_{\rho^{(\prime)} K^{*(\prime)\pm}} & = & \frac{C_{\rho^{(\prime)}}}{g_{\rho}} I_{cu}^{K^{*(\prime)}K\rho^{(\prime)}} \left( I_{11}^{K^*K\eta_u} + \sqrt{2} I_{11}^{K^*K\eta_s} \right) \left(Z_{K^{*(\prime)}K_1} + 1 \right) BW_{\rho^{(\prime)}}(q^2) BW_{K^{{*(\prime)}\pm}}(q_{\eta K^{\pm}}^2), \nonumber\\
    \mathcal{M}_{\omega^{(\prime)} K^{*(\prime)\pm}} & = & \frac{1}{3} \frac{C_{\omega^{(\prime)}}}{g_{\rho}} I_{cu}^{K^{*(\prime)}K\omega^{(\prime)}} \left( I_{11}^{K^*K\eta_u} + \sqrt{2} I_{11}^{K^*K\eta_s} \right) \left(Z_{K^{*(\prime)}K_1} + 1 \right) BW_{\omega^{(\prime)}}(q^2) BW_{K^{{*(\prime)}\pm}}(q_{\eta K^{\pm}}^2), \nonumber\\
    \mathcal{M}_{\phi^{(\prime)} K^{*(\prime)\pm}} & = & - \frac{2}{3} \frac{C_{\phi^{(\prime)}}}{g_{\phi}} I_{cs}^{K^{*(\prime)}K\phi^{(\prime)}} \left( I_{11}^{K^*K\eta_u} + \sqrt{2} I_{11}^{K^*K\eta_s} \right)
    \left(Z_{K^{*(\prime)}K_1} + 1 \right) BW_{\phi^{(\prime)}}(q^2) BW_{K^{{*(\prime)}\pm}}(q_{\eta K^{\pm}}^2), \nonumber\\
    \mathcal{M}_{\rho^{(\prime)}\rho^{(\prime)}} & = & 4 \frac{C_{\rho^{(\prime)}}}{g_{\rho}} m_u I_{30}^{\rho^{(\prime)} \rho^{(\prime)} \pi} I_{11}^{\rho^{(\prime)} K K} Z_{\rho^{(\prime)}}  BW_{\rho^{(\prime)}}(q^2) BW_{\rho^{(\prime)}}(q_{KK}^2), \nonumber\\
    \mathcal{M}_{\omega^{(\prime)}\omega^{(\prime)}} & = & \frac{4}{3} \frac{C_{\omega^{(\prime)}}}{g_{\rho}} m_u I_{30}^{\omega^{(\prime)} \omega^{(\prime)} \eta_u} I_{11}^{\omega^{(\prime)} K K} Z_{\omega^{(\prime)}}  BW_{\omega^{(\prime)}}(q^2) BW_{\omega^{(\prime)}}(q_{KK}^2), \nonumber\\
    \mathcal{M}_{\phi^{(\prime)}\phi^{(\prime)}} & = & \frac{8\sqrt{2}}{3} \frac{C_{\phi^{(\prime)}}}{g_\phi} m_s I_{03}^{\phi^{(\prime)} \phi^{(\prime)} \eta_s} I_{11}^{\phi^{(\prime)} K K} Z_{\phi^{(\prime)}}  BW_{\phi^{(\prime)}}(q^2) BW_{\phi^{(\prime)}}(q_{KK}^2),\nonumber\\
    \mathcal{M}_{\rho^{(\prime)} box} & = & \frac{C_{\rho^{(\prime)}}}{g_{\rho}} \left( I_{cu1box}^{\rho^{(\prime)}KK\eta_u} - 2\sqrt{2} I_{cu2box}^{KK\eta_s}\right) BW_{\rho^{(')}}(q^2), \nonumber\\
     \mathcal{M}_{\omega^{(\prime)} box} & = & \frac{1}{3} \frac{C_{\omega^{(\prime)}}}{g_{\rho}} I_{cubox}^{\omega^{(\prime)}KK\pi} \left( I_{cu1box}^{\rho^{(\prime)}KK\eta_u} - 2\sqrt{2} I_{cu2box}^{KK\eta_s}\right) BW_{\omega^{(\prime)}}(q^2), \nonumber\\
     \mathcal{M}_{\phi^{(\prime)} box} & = & \frac{2}{3} \frac{C_{\phi^{(\prime)}}}{g_{\phi}} \left( I_{cu2box}^{\rho^{(\prime)}KK\eta_u} - \sqrt{2} I_{cu2box}^{KK\eta_s}\right) BW_{\phi^{(\prime)}}(q^2), 
\end{eqnarray}
where the constants $Z$ arising after taking into account nondiagonal axial-vector and pseudoscalar transitions are defined in (\ref{avp_const}). In this case, the constant $Z_{K^{*(\prime)}K_1}$ has a different form and in addition $Z_{\phi^{(\prime)}}$ appears due to the presence of $\phi^ mesons {(\prime)}$ as the second intermediate states
\begin{eqnarray}
    Z_{K^{*(\prime)}K_1} & = & 1 - \frac{m_s I_{11}^{K^{*(\prime)}K_1\eta_u} + \sqrt{2} m_u I_{11}^{K^{*(\prime)}K_1\eta_s}}{I_{11}^{K^{*(\prime)}K\eta_u}+\sqrt{2}I_{11}^{K^{*(\prime)}K\eta_s}} I_{11}^{K_1K} \frac{m_s + m_u}{M_{K_{1A}}^2}, \nonumber\\
    Z_{\phi^{(\prime)}} & = & 1 - \frac{I_{11}^{\phi^{(\prime)} K_1 K}I_{11}^{K_1K}}{I_{11}^{\phi^{(\prime)} K K}} \frac{\left(m_u + m_s\right)^2}{M_{K_{1A}}^2}.   
\end{eqnarray}

Convergent integrals arises in diagrams with $VVP$ vertices of an anomalous type and box diagrams with $VPPP$ vertices. The integrals $I_{cu1box}=I_{cubox}$, $I_{cs1box}=I_{csbox}$ are defined in (\ref{int2}). The convergent integrals $I_{cu2box}$ and $I_{cs2box}$ arise after taking into account diagrams with the $s$ quark part of the $\eta$ meson
\begin{eqnarray}
    I_{cu2box}^{mesons} & = & m_u I_{22}^{mesons} + m^2_u\left(m_s - m_u\right)I_{32}^{mesons}, \nonumber\\
    I_{cs2box}^{mesons} & = & 2m_s I_{13}^{mesons} + \frac{m_s-m_u}{2} m^2_u I_{23}^{mesons} - \frac{3(m_s-m_u)}{2} m^2_s I_{14}^{mesons}.
\end{eqnarray}

\subsection*{Acknowledgments}
The authors thank prof. A. B. Arbuzov for useful discussions.

\end{document}